%% file: ICC_2015_v6.tex
\documentclass[10pt, conference, letterpaper]{IEEEtran}

\usepackage{cite}
\usepackage{graphicx,color,epsfig,rotating}
\usepackage{amsfonts,amsmath,amssymb,bbm}
\usepackage{algorithm,algorithmic}
\usepackage{subfigure}
\usepackage{amsmath}
\usepackage{cite}
\usepackage{mdwtab}
\usepackage{placeins}
\usepackage{psfrag, graphicx}
\usepackage[latin1]{inputenc}
\usepackage{amssymb}
\usepackage{multirow}
\usepackage{mathrsfs}
\input{macros}

\newtheorem{defn}{Definition}%[section]
\newtheorem{example}{Example}%[section]
\newtheorem{theorem}{Theorem}%[section]
\newtheorem{corollary}{Corollary}%[section]
\begin{document}

\sloppy

%% Paper Title
%% You can use linebreaks \\ within to get better formatting as
%% desired.

%\title{Caching-Aided Coded Multicast for Efficient Content Delivery with Popularity Distribution}
%\title{An Efficient Coloring Algorithm for Caching-Aided Coded Multicasting}
\title{An Efficient Multiple-Groupcast Coded
Multicasting Scheme for Finite Fractional Caching}

%% Author names and affiliations:
%%
%% Avoiding spaces at the end of the author lines is not a problem with
%% conference papers because we don't use \thanks or \IEEEmembership.
%%
%% For several authors with only one affiliation:
%%
% \author{
%   \IEEEauthorblockN{Hui-Ting Chang and Stefan M.~Moser}
%   \IEEEauthorblockA{Department of Electrical and Computer Engineering\\
%     National Chiao Tung University (NCTU)\\
%     Hsinchu, Taiwan\\
%     Email: \{email-of-hui-ting,email-of-stefan\}@ieee.org}
% }
%%
%% For up to three affiliations:
%%
%%
%% For over three affiliations, or if they all won't fit within the width
%% of the page, use this alternative format:
%%
% \author{
%   \IEEEauthorblockN{
%     Giuseppe ** \IEEEauthorrefmark{1},
%     Mingyue Ji \IEEEauthorrefmark{2},
%     Antonia Tulino \IEEEauthorrefmark{3},
%     Jaime Llorca \IEEEauthorrefmark{3} and
%     Paola ** \IEEEauthorrefmark{2}}
%   \IEEEauthorblockA{
%     \IEEEauthorrefmark{1}School of Electrical and Computer Engineering\\
%     Georgia Institute of Technology, Atlanta, Georgia 30332--0250\\
%     Email: see http://www.michaelshell.org/contact.html}
%   \IEEEauthorblockA{
%     \IEEEauthorrefmark{2}A\rm lcatel-Lucent, Bell Labs, USA}
%   \IEEEauthorblockA{
%     \IEEEauthorrefmark{3}Starfleet Academy, San Francisco, California 96678-2391\\
%     Telephone: (800) 555--1212, Fax: (888) 555--1212}
%   \IEEEauthorblockA{
%     \IEEEauthorrefmark{4}Tyrell Inc., 123 Replicant Street, Los Angeles, California 90210--4321}
% }

 \author{
   \IEEEauthorblockN{
     Mingyue Ji\IEEEauthorrefmark{1},
     Karthikeyan Shanmugam\IEEEauthorrefmark{5},
       Giuseppe Vettigli\IEEEauthorrefmark{3},
     Jaime Llorca\IEEEauthorrefmark{2}, 
      Antonia M. Tulino\IEEEauthorrefmark{2}\IEEEauthorrefmark{3},
      Giuseppe Caire\IEEEauthorrefmark{1}\IEEEauthorrefmark{4} 
   %  Eldon Tyrell\IEEEauthorrefmark{4}
    }
     \IEEEauthorblockA{
     \IEEEauthorrefmark{2}A\rm lcatel Lucent, Bell labs, NJ.
     Email: \{jaime.llorca, a.tulino\}@a\rm lcatel-lucent.com}
   \IEEEauthorblockA{
     \IEEEauthorrefmark{1}EE Department, University of Southern California, CA.
     Email: mingyuej@usc.edu}
          \IEEEauthorblockA{
     \IEEEauthorrefmark{5}ECE Department, University of Texas, TX. Email: karthiksh@utexas.edu}
        \IEEEauthorblockA{
     \IEEEauthorrefmark{3}EE Department, Universit\'a di Napoli Federico II, Italy.  }
     \IEEEauthorblockA{
     \IEEEauthorrefmark{4}EECS Department, Technische Universit\"at Berlin, Germany. Email: caire@tu-berlin.de}
%     Telephone: (800) 555--1212, Fax: (888) 555--1212}
%   \IEEEauthorblockA{
%     \IEEEauthorrefmark{4}Tyrell Inc., 123 Replicant Street, Los Angeles, California 90210--4321}
 }

%% Use for special paper notices

%% To balance the two columns, you should reduce the text-height of
%% the last page using the following command:
%%%%%%%%%%%%%%%%%%%%%%%%%%%%%%%%%%%%%%%%%%%%%%%%%%%%%%%%%%%%%%%%%%%%%
%\addtolength{\textheight}{-9.35cm}
%%%%%%%%%%%%%%%%%%%%%%%%%%%%%%%%%%%%%%%%%%%%%%%%%%%%%%%%%%%%%%%%%%%%%
%% with an appropriate value. This command must be place on the second
%% last page, i.e., for a one-page abstract here, for a two-page
%% abstract right after the \maketitle command.

%% Create the title:
\maketitle

\begin{abstract}

Coded multicasting has been shown to improve the caching performance of content delivery networks with multiple caches downstream of a common multicast link. However, the schemes that have been shown to achieve order-optimal performance require content items to be partitioned into a number of packets that grows exponentially with the number of users \cite{shanmugam2014}. In this paper, we first extend the analysis of the achievable scheme in \cite{ji2014order2} to the case of heterogeneous cache sizes and demand distributions, providing an achievable scheme and an upper bound on the limiting average performance when the number of packets goes to infinity while the remaining system parameters are kept constant. We then show how the scheme achieving this upper bound can very quickly loose its multiplicative caching gain for finite content packetization. To overcome this limitation, we design a novel polynomial-time algorithm based on greedy local graph-coloring that, while keeping the same content packetization, recovers a significant part of the multiplicative caching gain. Our results show that the achievable schemes proposed to date to quantify the limiting performance, must be properly designed for practical finite system parameters.

\end{abstract}

\section{Introduction}
\label{section: intro}

Recent studies \cite{llorcatulino132, ji2013optimalJ, ji2014average, ji2015random, ji2014order2, maddah2012fundamental, maddah2013decentralized, ji2013wireless} have been able to characterize the information theoretic limiting performance of several caching networks of practical relevance, in which network load scales inversely linear with cache size, showing great promise to accommodate the exponential traffic growth experienced in today's communication networks \cite{cisco13}.

%Consider a network with one source (base station), having access to $m$ files, and $n$ users (caches), each with a storage capacity of $M$ files.
%In \cite{ji2013optimalJ}, 
%the authors showed that if the users can communicate between each other via Device-to-Device (D2D) communications, 
%a simple distributed random caching placement scheme and TDMA-based unicast D2D delivery achieves the order-optimal\footnote{Order-optimal means that the gap between the information theoretic converse and the achievable throughput can be bounded by a constant number when $m,n \rightarrow \infty$.} throughput $\Theta\left(\max\{\frac{M}{m}, \frac{1}{m}, \frac{1}{n}\}\right)$,\footnote{Given two functions $f$ and $g$, we say that: 1)  $f(n) = O\left(g(n)\right)$ if there exists a constant $c$ and integer $N$ such that  $f(n)\leq cg(n)$ for $n>N$ 2) $f(n) = \Theta\left(g(n)\right)$ if $f(n) = O\left(g(n)\right)$ and $g(n) = O\left(f(n)\right)$.}  
%whose linear scaling with $M$ when $Mn \geq m$ shows a remarkable caching potential. Moreover, in this scheme each user caches entire files without the need of partitioning files into packets, 
%and missed files are delivered via unicast transmissions from neighbor nodes, making it practically implementable in real scenarios. 

%In the case that users cannot communicate between each other, but 
We consider a network with $n$ users, each with a cache of size $M$ files, sharing a multicast link from a content source with access to a library of $m$ files. 
For the worst-case demand setting, in which each user places a distinct file request, 
the authors in \cite{maddah2012fundamental} presented a deterministic caching and coded multicasting scheme achieving the order-optimal \emph{transmission rate}~\footnote{We define transmission rate in terms of number of file transmissions or  number of file-unit capacity channel-uses.} 
$\Theta\left(\min\{\frac{m}{M}, {m}, {n}\}\right)$.\footnote{Given two functions $f$ and $g$, we say that: 1)  $f(n) = O\left(g(n)\right)$ if there exists a constant $c$ and integer $N$ such that  $f(n)\leq cg(n)$ for $n>N$ 2) $f(n) = \Theta\left(g(n)\right)$ if $f(n) = O\left(g(n)\right)$ and $g(n) = O\left(f(n)\right)$.} However, the scheme in \cite{maddah2012fundamental} 
requires a centralized caching policy and each file to be partitioned into a number of packets that grows exponentially with the number of users. In \cite{maddah2013decentralized}, the authors presented an alternative scheme for the same network that uses a simpler decentralized random caching policy while a more complex coded multicasting scheme requiring a number of computations that grows exponentially with the number of users. Nonetheless, to guarantee the same rate, the file size (or equivalently the number of packets per file) is required to go to infinity. In \cite{ji2014order2}, the authors extended the analysis to the case in which each user places $L \in \{1, \dots, m\}$ simultaneous file requests and provided an order-optimal delivery scheme based on local graph coloring~\cite{shanmugam2013local} that is able to optimally combine the gains from coded and naive multicasting
%~\footnote{We refer to naive multicasting as the transmission of a single uncoded file unit to multiple destinations.} 
depending on the level of overlapping created by the multiple per-user requests. %These results are again asymptotic in the number of packets per file.
% balance the gains from codedshowed that the order optimal ratethe local coloring scheme  can achieve order-optimality for the multiple requests scenario although it is not clear whether coloring based schemes are sufficient to achieve order-optimality. 
In \cite{ji2014average}, the authors considered the case in which user demands are characterized by a popularity distribution, and proposed a scheme consisting of a random popularity-based (RAP) caching policy and a chromatic-number index coding (CIC) multicasting scheme, referred to as RAP-CIC, proved to be order-optimal in terms of average rate.  
In order to analytically quantify the performance of RAP-CIC,  the authors in \cite{ji2014average} resorted to a polynomial-time approximation of CIC, referred as greedy constrained coloring (GCC) that guarantees the order-optimal rate in the asymptotic regime of infinite packetization. %RAP-GCC is shown to include and generalize the order-optimal scheme for the worst-case demand in \cite{maddah2013decentralized}. 
Using RAP-GCC, the authors further provided the regions of the system parameters, characterized by low popularity skewness and large aggregate cache size $nM>m$, in which multiplicative caching gains are potentially achievable.

\begin{figure}[ht]
\vspace{-0.3cm}
\centerline{\includegraphics[width=9cm]{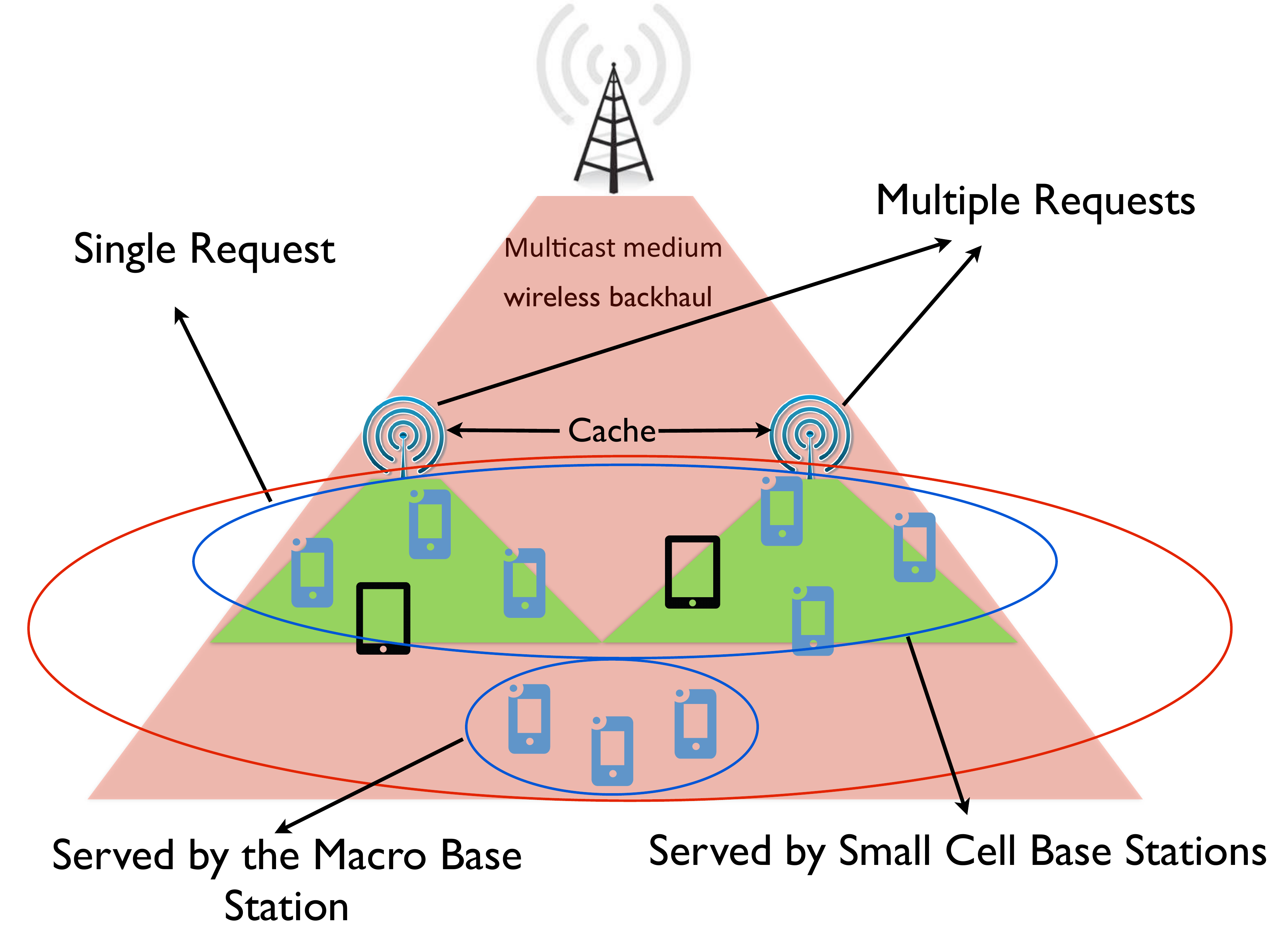}}
%\vspace{-1cm}
\vspace{-0.5cm}
\caption{An example of the network model, which consists of a source node (base station in this figure) having access to the content library and connected to the users via a shared bottleneck (multicast) link. Each user (end users and small cell base stations) may have different cache size and request a different number of files according to their own demand distribution.}
\label{fig: network_overview}
\vspace{-0.4cm}
\end{figure}

It is then of key importance to understand if using any of above mentioned schemes, the promising multiplicative caching gain can be preserved in practical settings with finite file packetization.   
In this paper, we try to address this question focusing  on a non-homogenous caching network with a shared multicast link, where users make $L \in \{1, \dots, m\}$ requests according to possibly different demand distributions and have possibly different cache sizes. % (see Fig.~\ref{fig: network_overview} as an example). 
As shown in Fig. \ref{fig: network_overview}, this scenario can be motivated by the presence of both user caches and cache-enabled small cell base stations, each of which serving a set of users. In this case, each small cell base station can be modeled as a user cache placing multiple requests. 
%The considered network is a generalization of the one considered in \cite{ji2014average},  where all the users are assumed to have the same demand distribution, equal storage capacity and one request per user. The contributions of this paper are as follows.
To this end, we first introduce RAP-GCLC (RAP-greedy constrainted local coloring) as an extension to RAP-GCC in order to quantify the average performance of the to above non-homogenous shared link network.
%To this end, we first extend RAP-CIC and RAP-GCC to the above non-homogenous shared link network, where users may have different storage capacities, different demand distributions, and different number of requests, and quantify their average performance for infinite packetization by introducing the delivery scheme referred to as greedy constraint local coloring (GCLC). 
%It is worthy to mention that the extension to the multiple requests is of importance. As shown in Fig. \ref{fig: network_overview}, the users can be modeled as small cell base stations, each of which serves a set of users. In this case, each small cell base station can be modeled as a user with multiple requests. 
Next, we focus on the regime of finite file packetization and numerically show that RAP-GCLC cannot  guarantee the asymptotic order-optimal performance. Consequently, we introduce a novel algorithm referred to as RAP-HgLC (RAP-hierarchical greedy local coloring), %based on a greedy
%randomized approach referred to as Greedy Randomized Algorithm Search
%Procedure (GRASP), 
which is shown to recover a significant part of the asymptotic multiplicative caching gain. This algorithm's running time is quadratic in the number of packets. %while incurring a complexity at most quadratic in the number of packets. 

The paper is organized as follows. Section \ref{section: network model} introduces the network model and problem formulation. The achievable caching and coded delivery scheme, along with the general upper bound on the average achievable rate are presented in Section \ref{sec: Achievable Delivery Scheme}. Section \ref{sec: algorithms} describes
the proposed polynomial-time delivery scheme. Finally, Section \ref{sec: Simulations and Discussions} presents the simulation results and related discussions. %, and we conclude the paper in Section \ref{sec: Conclusion}.

%%%%%%%%%%%%%%%%%%%%%%%%%%%%%%%%%%%%%%%%%%%%%%%%%%%%%%
%\vspace{-0.2cm}
\section{Network Model and Problem Formulation}
\label{section: network model}

We consider a network consisting of a source node with access to a content library $\Fc=\{1,\dots, m\}$  of files with size $F$ bits, and
 $n$ user nodes $\Uc=\{1,\dots, n\}$. We assume the source node communicates to the user nodes through a shared
multicast link of finite capacity $C$. Without loss of generality, we can assume $C = F$ bits/unit time
and measure the transmission rate of the scheme in units of  time necessary to deliver the requested messages to the users.
User $u \in \Uc$ has a storage capacity of size $M_uF$ bits (i.e., $M_u$ files).
The channel between the source and all the users follows a shared error-free deterministic model.
User $u$ makes $L_u$ file requests, each of which follows a probability distribution $q_{f,u}$, where $q_{f,u} \in [0,1]$ and $\sum_{f=1}^m q_{f,u} = 1$ (i.e., for each of the $L_u$ requests of user $u$, file $f$ is chosen with probability $q_{f,u}$). 
All the file requests (by one user or across users) are assumed to be placed independent of each other. %Without loss of generality, let $L_1 \geq L_2 \geq L_3, \dots, L_n$. 
We denote $\Qm = [q_{f,u}], \,u\in\{1, \dots, n\}, f\in\{1, \dots, m\}$, as the demand distribution. Let the requested files by user $u$ be $\fv_{u} = [f_{1, u}, f_{2, u}, \dots, f_{L_u, u}]$. %and $\Fm = \{\fv_u, u = 1, \dots, n\}$. 
%One example of the network model is shown in Fig. \ref{fig: network_overview}. 
The goal is to design a content distribution scheme (i.e., determine the information stored in the user caches and
the multicasted codeword to be sent to all users through the shared link) such that all demands are satisfied with probability $1$ and the
expected rate $R(\Qm)$ is minimized.\footnote{The expected rate is defined as the average minimum number of file transmissions.%, which is inversely proportional to the average throughput
}  The expectation is over the demand distribution $\Qm $. We denote the minimum achievable expected rate by $R^*(\Qm)$.

%%%%%%%%%%%%%%%%%%%%%%%%%%%%%%%%%%%%%%%%%%%%%%%%%%%%%%%%%%%
\section{Achievable Scheme}
\label{sec: Achievable Delivery Scheme}

In this section, we present an achievable scheme based on random popularity-based caching and index coding based delivery.

%%%%%%%%%%%%%%%%%%%%%%%%%%%%%%%%%%%%%%%%%%%%%%%%%%%
%\vspace{-0.1cm}
\subsection{Caching Placement}
\label{sec: Caching Placement Scheme}

We partition each file into $B$ equal-size packets, represented as symbols of $\FF_{2^{F/B}}$, where $F/B$ is sufficiently large (see later).
Let $\Cm$ and $\Wm$ denote the realizations of the {\em packet level}
caching and demand configurations, respectively,
where $\Cm_{u,f}$ denotes the packets of file $f$ cached
at user $u$, and $\Wm_{u,f}$ denotes the packets of file $f$ requested by user $u$. %We use the caching algorithm shown in Fig.~\ref{alg1} 
%to 
We let each user fill its cache independently (and therefore in a decentralized way) by knowing the 
caching distribution $\Pm=[p_{f,u}]$, $u\in\{1, \dots, n\}, f\in\{1, \dots, m\}$, with $\sum_{f=1}^m p_{f,u}=1, \forall u$ and $0 \leq p_{f,u}\leq1/M_u, \forall f$. The caching placement is shown in Algorithm \ref{alg1}. 

\begin{algorithm}
\caption{Random Popularity-Based Caching (RAP)}
\label{alg1}
{\small
\begin{algorithmic}[1]
%\REQUIRE $\Pm=p_{f,u}, f=1, \dots, m, u = 1, \dots, n$
\FORALL{$f \in \mathcal{F}$}
\STATE Each user $u$ caches a subset ($\Cm_{u,f}$) of $p_{f,u} M_u B$ distinct packets of file $f$ uniformly at random. %for all the users.
%\STATE Each user $u$ caches $p_{f,u} M B$ packets independently of file $f$ according to the caching distribution $p_{f,u}$.
\ENDFOR
\RETURN $\Cm = [\Cm_{u,f}], u\in\{1, \dots, n\}, f\in\{1, \dots, m\}$.
\end{algorithmic}
}
\end{algorithm}

%\begin{figure}[th!]
%\centerline{\fbox{\begin{minipage}[h]{7 cm} \normalsize{
%{\bf algorithm} {\tt Caching algorithm} \\
%$1$\hspace*{0.2truecm} {\bf for} $f \in \mathcal{F}$ $\rightarrow$\\
%%$2$\hspace*{0.7truecm} Each user $u$ randomly caches a subset of \\
%%\hspace*{0.8truecm} packets with cardinality of $p_{f,u} M B$  \\
%%\hspace*{0.8truecm}  from file $f$ according to $[p_{f,u}]_{f=1}^m$; \\
%$2$\hspace*{0.7truecm} Each user $u$ caches a subset of \\
%\hspace*{0.8truecm} $p_{f,u} M_u B$ distinct packets of file $f$ \\
%\hspace*{0.8truecm}  uniformly at random; \\
%$3$\hspace*{0.1truecm} {\bf endfor} \\
%$4$\hspace*{0.5truecm} $\Cm = \{\Cm_{u,f}, u = 1, \dots, n, $ \\
%$\hspace*{2truecm} f = 1, \cdots, m\}$; \\
%$5$\hspace*{0.0truecm} {\bf return}($\Cm$);          \\
%{\bf end} {\tt Caching algorithm} }
%\end{minipage}}}
%\caption{The distributed random caching algorithm.}
%\label{alg1}
%\end{figure}
%%%%%%%%%%%%%%%%%%%%%%%%%%%%%%%%%%%%%%%%%%%%%%%%%%%%%

%%%%%%%%%%%%%%%%%%%%%%%%%%%%%%%%%%%%%%%%%%%%%%%
\subsection{Coded Multicast Delivery}
\label{sec: Coded Transmission}

Our coded delivery scheme is based on local chromatic number index coding \cite{shanmugam2013local, ji2014order2}. %\cite{blasiak2010index, ji2014average}.
The directed %(undirected) 
conflict graph $\mathcal H_{\Cm, \Wm} = (\Vc, \Ec)$ %, where $\Vc$ and $\Ec$ denote the set of vertices and edges of $\mathcal H_{\Cm, \Wm}$, respectively, 
is constructed as follows:
\begin{itemize}
\item Consider each packet requested by each user as a distinct vertex in %, i.e., if the same packet is requested by $N > 1$ users, it results in $N$ distinct vertices. Hence, each vertex of 
$\mathcal H_{\Cm, \Wm}=(\Vc, \Ec)$. Hence, each vertex $v\in\Vc$ is uniquely  identified by the pair 
$\{ \rho(v),\mu(v)\}$  where $\rho(v)$ indicates the  \mbox{packet identity} associated to the vertex and 
$\mu(v)$ represents the \mbox{user requesting it}.
\item For any pair of vertices $v_1$, $v_2$, we say that vertex (packet) $v_1$ interferes with vertex $v_2$ 
if the packet associated to the vertex $v_1$, $\rho(v_1)$, is not  in the cache of the user associated to vertex  $v_2$, $\mu(v_2)$, 
and $\rho(v_1)$ and $\rho(v_2)$ do not represent the same packet. Then, 
draw a directed edge from vertex $v_2$ to vertex $v_1$ if $v_1$ interferes with $v_2$.
%Create an edge between vertices $v_1, v_2 \in \Vc$ if: 1) they do not represent the same packet,
%and 2) $v_1$ is not available in the cache of the user requesting $v_2$,
%or $v_2$ is not available in the cache of the user requesting $v_1$.
\end{itemize}

%Based on this caching scheme, we design a delivery scheme based on linear index coding.  
%(i.e., the encoding function is a linear function over an extension field of $\FF_2$). 
We focus on encoding functions of the following form: for the request vectors $\fv_{u}, u \in \Uc$, the multicast codeword is given by
\be  \label{VV}
X_{\{\fv_{u}, u \in \Uc\}} = \sum_{v \in \mathcal{V}} \omega_v \gv_v = \Gm \omegav,
\ee
where %$\mathcal{O}$ is the set of requested sub-packets, 
$\omega_v$ is the binary vector corresponding to packet $v$, represented as a (scalar) symbol of
the extension field $\FF_{2^{F/B}}$, the $\nu$-dimensional vector $\gv_v \in \FF_{2^{F/B}}^\nu$ is the coding vector of packet $\rho(v)$ and where
we let $\Gm = [\gv_1, \dots \gv_{|\mathcal{V}|}]$ and $\omegav= [\omega_1 ,\dots, \omega_{|\mathcal{V}|}]^{\transp}$. 
The number of { columns } $\nu$ of $\Gm$ yields the number of packet transmissions. Hence, the 
transmission rate is given by $\nu/B$ file units.  
%For the delivery scheme, 
To find the desired $\nu$, we introduce the definition of the local chromatic number:
\begin{defn} ({\bf Local Chromatic Number})
The directed local chromatic number of a directed graph $\mathcal{H}^d$ is defined as: 
\begin{align}
& \chi_{\rm lc}(\mathcal H^d) = \min_{\cv \in \mathcal{C}} \max_{v\in\mathcal{V}}|\cv(\mathcal{N}^+(v))|
\end{align}
where $\mathcal{C}$ denotes the set of all vertex-colorings of $\mathcal{H}$,  with  $\mathcal{H}$ indicating the  undirected version of $\mathcal H^d$,\footnote{An edge is undirected if the edge is present in either direction.} 
$\mathcal{V}$ denotes the vertices of $\mathcal{H}^d$, $\mathcal{N}^+(v)$ is the closed out-neighborhood of 
vertex $v$,\footnote{Closed out-neighborhood of vertex $v$ includes vertex $v$ and all the connected vertices 
via out-going edges of $v$.} 
and $\cv(\mathcal{N}^+(v))$ is the total number of colors in $\mathcal{N}^+(v)$ for the given coloring $c$. 
\hfill $\lozenge$
\end{defn}

It can be shown that, for sufficiently large $F/B$, there exists a $\Gm$ such that a valid index code can be found whose length is equal to the local chromatic number, and whose correspoding transmission rate is given by
$\chi_{\rm lc}(\mathcal H^d)/B$. We refer to this coding scheme as LCIC (local chromatic index coding).
The design of $\Gm$ is given by \cite{shanmugam2013local, ji2014order2}. \footnote{Instead of using local chromatic number it is also straightforward to use fractional local chromatic number to design the coding vector $\Gc$ as illustrated in \cite{shanmugam2013local , ji2014order2}.}, and an example is given in the following: %Example \ref{example: algorithm} in the following for illustration.  

\begin{example}
\label{example: algorithm}
We consider a network with $n=3$ users denoted as $\mathcal U = \{1,2,3\}$ and $m=3$ files denoted as $\mathcal F = \{\mathrm{A}, \mathrm{B},\mathrm{C}\}$. We assume $M=1$ and sub-packetize each file into three packets. For example, $\mathrm{A} = \{\mathrm{A}_1, \mathrm{A}_2, \mathrm{A}_3\}$. Let $p_{\mathrm{A},u} = \frac{1}{3}$, $p_{\mathrm{B},u} = \frac{1}{3}$ and $p_{\mathrm{C},u} = \frac{1}{3}$ for $u \in \{1,2,3\}$, which means that one packet from each of $\mathrm{A}, \mathrm{B}, \mathrm{C}$ will be stored in each user's cache. We assume a caching realization $\Cm$ is given by: user $u$ caches $\{\mathrm{A}_u,\mathrm{B}_u,\mathrm{C}_u\}$ ($\Cm_{u,\mathrm{A}} = \{\mathrm{A}_u\}, \Cm_{u,\mathrm{B}} = \{\mathrm{B}_u\}, \Cm_{u,\mathrm{C}} = \{\mathrm{C}_u\}$). We let each user make one request. Specifically, we let
user $1$ request $\mathrm{A}$, user $2$ request $\mathrm{A}$ and user $3$ request $\mathrm{B}$ ($\fv_1 = \{\mathrm{A}\}, \fv_2=\{\mathrm{A}\}, \fv_3 = \{\mathrm{B}\}$) such that $\Wm_{1,A} = \{\mathrm{A}_{2},\mathrm{A}_3\}, \Wm_{2,A} =\{\mathrm{A}_{1}, \mathrm{A}_{3}\}, \Wm_{3,B} = \{\mathrm{B}_1,\mathrm{B}_2\}$. The conflict graph and the corresponding coloring are shown in Fig. \ref{fig: conflict_ex_coloring}. We can see that the total number of colors needed, the chromatic number in this case, is $5$, while the local coloring number, or the local chromatic number in this case, is $4$. We construct $\Gm$ by using the generator matrix of a $(5,4)$ MDS code, which is given by: 
\be
%\Vm' = 
\left( \begin{array}{cccccc} 1 & 0 & 0 & 0 & 1 \\ 
0 & 1 & 0 & 0 & 1 \\
0 & 0 & 1 & 0 & 1 \\ 
0 & 0 & 0 & 1 & 1 \\
%0 & 0 & 0 & 0 & 1 & 1 \\
\end{array}\right).
\ee
Then, we allocate the same vector to the vertex (packet) with the same color as shown in Fig. \ref{fig: conflict_ex_coloring}. Hence, the transmitted codeword is given by 
$A_1 \oplus A_2$, $A_1 \oplus A_3$, $A_1 \oplus B_1$, $A_1 \oplus B_2 $, of length
$4/3$ file units.
\begin{figure}[ht]
%\vspace{-0.5cm}
\centerline{\includegraphics[width=7cm]{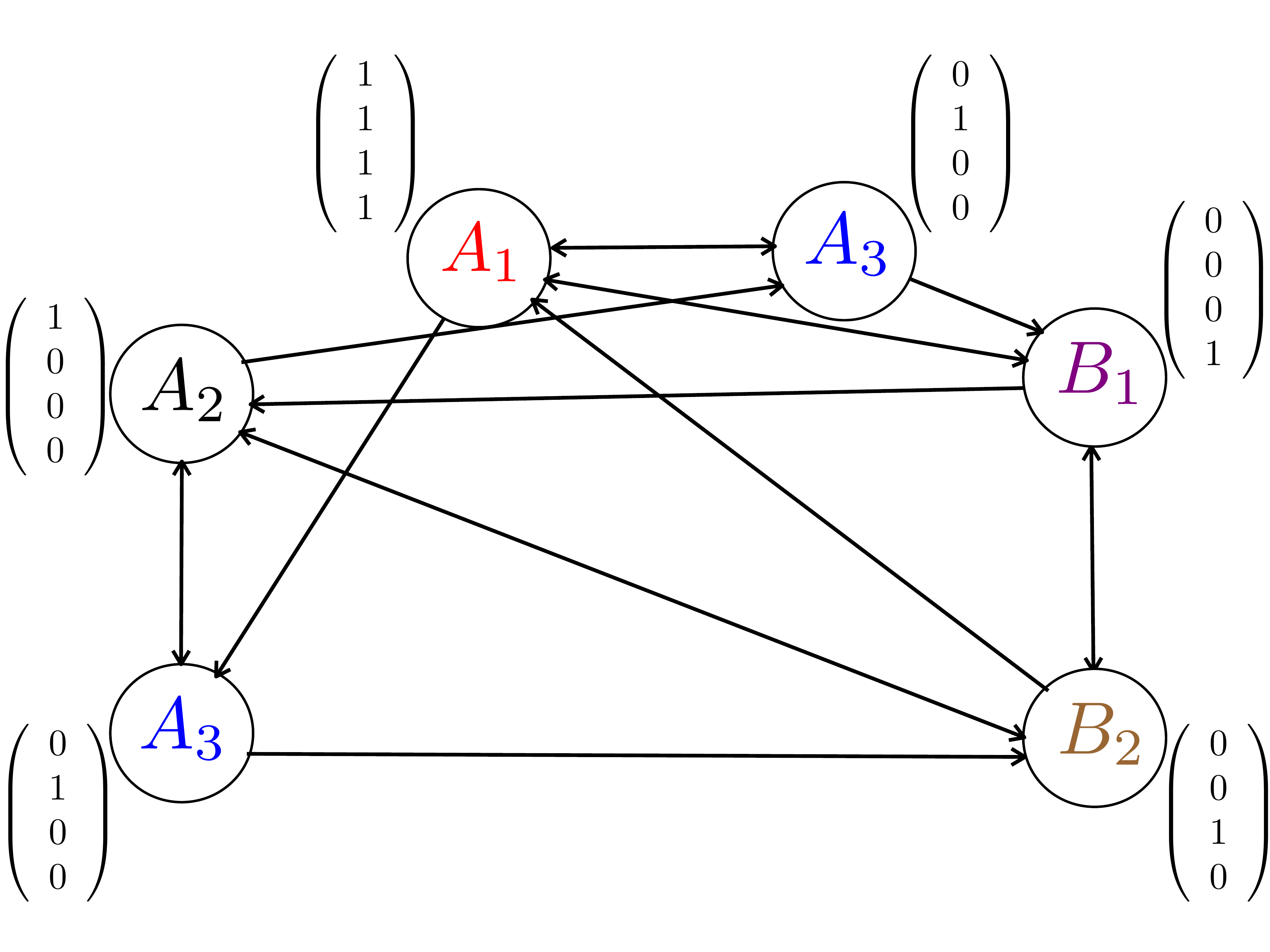}}
%\vspace{-0.5cm}
\caption{An illustration of the directed conflict graph and the corresponding index code. The coloring of the graph is given by the colors of the fonts. The total number of colors is $5$, and the local coloring number is $4$.} %where $n=3$, $\mathcal U = \{1,2,3\}$, $m=3$, $\mathcal F = \{\mathrm A, \mathrm B, \mathrm C\}$ and $M=1$. Each file is partitioned into $3$ packets. For example, $\mathrm A = \{\mathrm A_1, \mathrm A_2, \mathrm A_3\}$. The caching realization $\Cm$ is that user $1$ caches $\{\mathrm A_1, \mathrm A_2, \mathrm B_1\}$; user $2$ caches $\{\mathrm A_1, \mathrm A_3, \mathrm B_2\}$; user $3$ caches $\{\mathrm A_1, \mathrm A_2, \mathrm B_3\}$. The requests vector is $\fv = \{\mathrm A, \mathrm B, \mathrm C\}$.
%Hence, $\Wm = \{\mathrm A_3, \mathrm B_1, \mathrm B_3, \mathrm C_1, \mathrm C_2, \mathrm C_3\}$. The color for each vertex in this graph represents a vertex coloring scheme achieved by the algorithm given by Fig.~\ref{algorithm: coloring 1}. Note that in this case, this vertex coloring is the minimum vertex coloring and the resultant number of packet transmissions is $5$.}
%\vspace{-0.5cm}
\label{fig: conflict_ex_coloring}
\end{figure}
%\vspace{-0.3cm}
\hfill $\lozenge$
\end{example}

It is worth noticing that given $\Cm$ and $\Wm$, for any  valid coloring scheme $\bf c$ of $\mathcal H_{\Cm, \Wm}$,   and its associated local coloring number, for sufficiently large $F/B$, there always exists an
index code  $\Gm$, such that the total number of transmissions in terms of packets is given by
the local coloring number $\max_{v\in\mathcal{V}}|\cv(\mathcal{N}^+(v))|$, and the corresponding  transmission rate
is $\max_{v\in\mathcal{V}}|\cv(\mathcal{N}^+(v))|/B$. 
%A possible example of a valid coloring  is the one corresponding to the minimum number of colors on the graph or to any good approximation of it.
%Ideally, the minimum transmission rate given by local chromatic number is $\chi_{\rm lc}(\mathcal H_{\Cm, \Wm})/B$. In the following we refer to this coding scheme scheme as LCIC (local chromatic index coding). 
%Due to the difficulty of characterizing the exact value of $\chi_{\rm lc}(\mathcal H_{\Cm, \Wm})$, instead, in the following, we will capture an upper bound of $\chi_{\rm lc}(\mathcal H_{\Cm, \Wm})$. In other words, we will propose an efficient coloring scheme called Hierarchical greedy Local Coloring (HgLC) such that the achievable local coloring number or transmission rate achieves a significant gain compared with the conventional caching scheme in literature. 

\subsection{Achievable Expected Rate}

%{ \RED Mingyue:  here we have problem q is a matrix now not a vector any more}
Given $n,m,M$ and the demand distribution $\Qm$, our goal is to find the caching distribution $\Pm$ that minimizes the expected
rate $R^{\rm LCIC}(\Pm,\Qm) \eqdef  \lim_{B \rightarrow \infty }\EE[\chi_{\rm lc}(\Hsf_{\Csf, \Wsf})/B]$.\footnote{$\Hsf_{\Csf, \Wsf}$ denotes the random conflict graph, which is a function of the random caching and demand configurations, $\Csf$ and $\Wsf$, respectively.} Let $L = \max_u L_u$ and order $L_u, u \in \Uc$ as a decreasing sequence $L_{[1]} \geq L_{[2]} \geq L_{[3]}, \dots, L_{[n]}$, where $L_{[i]}$ is the $i$-th largest $L_u$ and $[i] = u$ for some $u \in \Uc$. It can be seen that $L_{[1]} = \max_u L_u$ and $L_{[n]} = \min_u L_u$. 
Let $n_j = \sum_{[i]} 1\{L_{[i]} - j \geq 0\}$, where $1 \leq j \leq {L_{[1]}}$  and $1\{\cdot\}$ is the indicator function. Let $\Uc_{n_j} = \{[i] \in \Uc: 1\{L_{[i]} - j \geq 0\}\}$. 
%We partition the set of the requests into $L = \max_u L_u$ sets denoted as $\Lc_1, \cdots, \Lc_{L}$, where the users requesting files in $\Lc_i$ are all distinct, each requested file by user $u, u \in \Uc$ only appears in one of $\Lc_i, i = 1, \cdots, L$, and $\cup_{i=1}^L \Lc_i = \cup_{u=1}^n \fv_u$. Let $\Uc(\Lc_i)$ denote the set of users requesting the files in $\Lc_i$. 
 In the next theorem we provide an upper bound of $R^{\rm LCIC}(\Pm,\Qm)$ 
given by the rate achievable with a greedy constrained local coloring (GCLC) scheme which is described in details in the next section (see Section IV-A) and can be seen as a generalization of the GCC scheme presented in  \cite{ji2014average}.
%The upper bound of is given by the following theorem:
\begin{theorem}
\label{thm:up}
For any given $m$, $n$, $M_u$, and $\Qm$, when $B \rightarrow \infty$, 
the expected rate $R^{\rm LCIC}(\Pm,\Qm)$ achieved by a content distribution scheme that uses caching policy in Algorithm~\ref{alg1} with caching distribution $\{\Pm = [p_{f,u}]: \sum_{f=1}^mp_{f,u} = 1, \forall u; 0 \leq p_{f,u}\leq1/M_{u},\forall f,u\}$, and LCIC transmission,
satisfies:\footnote{A stronger version Theorem  \ref{thm:up}  states the following stronger result:  
$
\displaystyle \lim_{F \rightarrow \infty} \PP \left(\mathbb{E}[\chi_{\rm lc}(\Hsf_{\Csf, \Wsf})/B] \leq R^{\rm GCLC}(\Pm,\Qm ) + \epsilon \right)   = 1$, where the expectation is taken over only the demand distribution $\Qm$.
 }
\begin{align} 
\label{eq:2}
R^{\rm LCIC}(\Pm,\Qm)
\leq  R^{\rm GCLC}(\Pm,\Qm) \eqdef \min\{\psi(\Pm,\Qm),\bar m - \bar M\},
\end{align}

In (\ref{eq:2}), 
\begin{eqnarray}
\label{eq: m bar}
\bar m=\sum_{f=1}^m \left(1 - \prod_{u=1}^n \left(1 - q_{f,u}\right)^{L_u} \right),
\end{eqnarray}
and
\be 
\bar M = \sum_{f=1}^m \min_{u} p_{f,u} \left(1 - \prod_{u=1}^n \left(1 - q_{f,u}\right)^{L_u} \right),
\ee 
and
\begin{eqnarray}
\label{eq: psi}
\psi(\Pm,\Qm) &=& \sum_{j = 1}^{L} \sum_{\ell=1}^n \sum_{\Uc^\ell \subset \Uc_{n_j}}  \sum_{f=1}^m \sum_{u \in \Uc^\ell} \notag\\
&& \rho_{f, u, \Uc^\ell} (1-p_{f,u} M_u)^{n_j-\ell+1} (p_{f,u} M_u)^{\ell-1},  \notag %\\
\end{eqnarray}
where $\Uc^\ell$ denotes a set of users with cardinality $\ell$ and
\begin{eqnarray}
&& \displaystyle 
\rho_{f, u, \Uc^\ell} \eqdef 
\notag\\
&& 
\mathbb \PP(f = \arg\! \max_{f_u \in \fv(\Uc^\ell)} \,\,\, (p_{f,u}M_u)^{\ell-1}(1-p_{f, u}M_u)^{n_j-\ell+1}), \notag%\\
\end{eqnarray}
denotes the probability that $f$ is the file whose $p_{f,u}$ maximizes the term $(p_{f, u}M_u)^{\ell-1}(1-p_{f, u}M_u)^{n-\ell+1})$ among $\fv(\Uc^\ell)$ (the set of files requested by $\Uc^\ell$).
\hfill  $\square$
\end{theorem}
%Theorem \ref{thm:up} is proved in Appendix \ref{sec: Proof of Theorem up}.

%\begin{example}
Under homogeneous demand distribution, cache size and number of request per user, %$q_{f,u} = q_f, M_u=M, L_u = L, \forall u \in \Uc$ and $L = \{1, \cdots, m\}$, then $p_{f,u} = p_f, \forall u \in \Uc$, 
we have the following corollary: %and (\ref{eq: m bar}) and (\ref{eq: psi}) become
%\begin{align}
%\label{eq:2 simplified}
%\bar R(\Pm,\Qm)
%\leq \bar R^{\rm ub}(\Pm,\Qm) \eqdef \min\{\psi(\Pm,\Qm),\bar m\},
%\end{align}
%where
\begin{corollary}
\label{corollary: result homogeneous}
Let  $q_{f,u} = q_f, M_u=M, L_u = L, \forall u \in \Uc$ and $L = \{1, \dots, m\}$, then $p_{f,u} = p_f, \forall u \in \Uc$ and when $B \rightarrow \infty$, $R^{\rm LCIC}(\Pm,\Qm)$ is given by (\ref{eq:2}), where 
\be
\label{eq: m bar simplified}
\bar m = \sum_{f=1}^m \left(1 - \left(1 - q_f\right)^{nL} \right),
\ee
and
\be 
\label{eq: M bar simplified}
\bar M = \sum_{f=1}^m p_{f} \left(1 - \left(1 - q_{f}\right)^{nL} \right),
\ee
and
\begin{eqnarray}
\label{eq: psi simplified}
&&\psi(\Pm,\Qm) = \notag\\
&&L\sum_{\ell=1}^n {n \choose \ell}  \sum_{f=1}^m \rho_{f,\ell} (1-p_f M)^{n-\ell+1} (p_f M)^{\ell-1},
\label{eq:3}
\end{eqnarray}
where
%\be
$\displaystyle \rho_{f,\ell} \eqdef \mathbb \PP(f = \arg\! \max_{j \in \Dc} \,\,\, (p_jM)^{\ell-1}(1-p_jM)^{n-\ell+1})$
%\ee
denotes the probability that file $f$ is the file whose $p_f$ maximizes the term $\left((p_jM)^{\ell-1}(1-p_jM)^{n-\ell+1}\right)$ among $\Dc$, which is a set of $\ell$ i.i.d. demands distributed as $\qv$. It can be seen that $ \rho_{f,\ell}$ is easy to evaluate. %among $\Fc^\ell$ (the set of files requested by an arbitrary subset of users of size $\ell$). 
%Eq. (\ref{eq: psi simplified}) is indeed the upper bound used in \cite{ji2014average} to obtain the order-optimal caching distribution for the homogenous network model. %Fig. ** shows the significant performance improvement that can be achieved
\hfill  $\square$
\end{corollary}
Due to space limitations, the proof of Theorem \ref{thm:up} is not included in this paper. Corollary \ref{corollary: result homogeneous} can be obtained directly from Theorem \ref{thm:up}. %, %but will prove Corollary \ref{corollary: result homogeneous} in Appendix \ref{sec: Proof of Corollary}. 

Using the explicit expression for $R^{\rm GCLC}(\Pm,\Qm )$ in Theorem \ref{thm:up}, we can optimize the 
caching distribution for a wide class of heterogeneous network models in order to minimize the number of transmissions. %In the following, we refer to the scheme that uses $\Pm^*$ for the caching algorithm in Fig~\ref{alg1} and CIC delivery as {\em Generalized RAndom Popularity-based} (GRAP), with achievable rate $\bar R(\Pm^*,\Qm)$.
We use $\Pm^*$ to denote the caching distribution that minimizes $R^{\rm GCLC}(\Pm,\Qm )$. It is worth noticing that for the homogeneous case described above, where $q_{f,u} = q_f, M_u=M, L_u = 1, \forall u \in \Uc$, $R^{\rm GCLC}(\Pm^*,\Qm)$ is indeed order optimal,  as proved in \cite{ji2014average}. 

%{\RED maybe fig***}

%In \cite{ji2014average}, it shows that RAP, \footnote{In \cite{ji2014average}, we refer the caching scheme with $\Pm^*$ together with the CIC-based coded multicast transmission as {\em RAndom Popularity-based (RAP)}} where $\Pm^*$ is obtained by optimizing (\ref{eq:2 simplified}), is order-optimal and can significantly reduce the average rate compared to other state-of-the-art caching and delivery scheme.
%\hfill $\lozenge$
%\end{example}

%\vspace{-0.2cm}

\section{Polynomial-time Algorithms}
\label{sec: algorithms}

%{ \RED maybe we can write differently In Section \ref{sec: Coded Transmission}, we can see that only the local %coloring number $\max_{v\in\mathcal{V}}|\cv(\mathcal{N}^+(v))|$ and the corresponding coloring $\cv$ are needed %for the coded multicasting delivery. Hence, 
In this section, we propose two efficient coloring algorithms for coded multicasting in heterogeneous shared link caching networks. 
We first introduce a polynomial-time greedy constrained local coloring (GCLC) algorithm, which generalizes the greedy constrained coloring (GCC) used in \cite{ji2014average} to quantify the order-optimal performance of homogeneous shared link networks in the asymptotic regime of $B \rightarrow \infty$. 
%This scheme is based on a greedy constrained local coloring (GCLC) approach, which is a generalization of the greedy constrained coloring (GCC) used in \cite{ji2014average}. %In Appendix \ref{sec: Proof of Theorem up}, 
In fact, GCLC is the scheme whose asymptotic ($B \rightarrow \infty$) average rate for heterogeneous shared link networks has been used in (\ref{eq:2}) to upper bound the asymptotic rate of LCIC. 
{ It is also easy to verify that GCLC achieves the same performance as the algorithm given in \cite{ji2014order2} for the worst-case demand setting in homogeneous shared link networks. }%In the following we will refer to this algorithm as GCC (Greedy Constrained Coloring).

We then present a novel coded multicasting algorithm called hierarchical greedy local coloring (HgLC) %based on a greedy random coloring approach 
that fully exploits the structure of the problem and also exhibits polynomial-time complexity. % than GCC and refer to it as GRASP (Greedy Randomized Algorithm Search Procedure). 
In Section \ref{sec: Simulations and Discussions}, we show that for finite file packetization, while GCLC loses the multiplicative caching gain, HgLC is able to approach the limiting performance and recover a significant part of the multiplicative caching gain.

\subsection{GCLC (Greedy Constrained Local Coloring)}
% An Algorithm achieving $\bar R^{\rm ub}(\Pm,\Qm)$
\label{sec: An Algorithm achieving upper bound}

%In this section, we recapitulate the GCC which has been introduced in \cite{ji2014average} and is a polynomial-time algorithm  achieving  $\bar R^{\rm ub}(\Pm,\Qm)$ when $F \rightarrow \infty$ and having the same performance as the exponentially (with $n$) complex algorithm in \cite{maddah2013decentralized}. 

The GCLC algorithm works by computing two valid local colorings of the conflict graph $\Hc_{\Cm,\Wm}$, referred to as GCLC$_1$ and GCLC$_2$. GCLC then compares the rate achieved by the two coloring solutions and constructs the transmission code based on the coloring with minimum rate. %\footnote{Recall that the transmission code (index code) is constructed by the modulo sum of all the vertices (packets) in $\Hc_{\Cm,\Wm}$ with the same color.}

Let $\Wm_u$ be  the set of requested packets by user $u$, and $\Cm_u$ the set of cached packets by user $u$. We define $\Tc_v = \{\mu(v)\} \cup \{u \in \Uc: \rho(v) \in  \Cm_u\}$.
%and   $\Kc_v = \{u \in \Uc: \rho(v) \in \Wm_{u} \cup \Cm_u\}$. 
Then, GCLC$_1$ is given by Algorithm \ref{algorithm: GCLC 1}. 
Observe that GCLC$_1$ computes a valid coloring of the conflict graph $\Hc_{\Cm,\Wm}$ and its associated local coloring number.  %as described in Fig.~\ref{algorithm: coloring 1}.  
Note that both the outer while-loop starting at line 3 and the inner for-loop starting at line 6 iterate at most $|\Vc|$ times, respectively, while all other operations inside the loops take constant time. %The operation in line 7 has complexity $O(n)$. 
Therefore, the complexity of GCLC$_1$ is $O(|\Vc|^2)$, %or equivalently $O(n^2B^2)$ since $|\Vc|\leq n B$, 
which is polynomial in $ |\Vc|$ (or $n,B$).

\begin{algorithm}[ht]
\caption{GCLC$_1$}
\label{algorithm: GCLC 1}
\small{
\begin{algorithmic}[1]
%\STATE Let ${\mathcal H}_{\Cm, \Wm} = (\Vc, \Ec)$ , where the set of vertices in ${\mathcal H}_{\Cm, \Wm}$ be $\Vc$
%\STATE Let $\hat \Vc = \Vc$; 
\STATE Let $\Cc = \emptyset$; 
\STATE Let $\cv = \emptyset$; 
\WHILE{$\Vc \neq \emptyset$}
\STATE Pick an arbitrary vertex $v$ in $\Vc$; Let $\Ic = \{v\}$;
\STATE Let $\Vc' = \Vc \setminus \{v\}$;
\FORALL{$v' \in \Vc'$ with $|\Tc_{v'}| = |\Tc_{v}|$}
\IF {\{There is no edge between $v'$ and $\Ic$\}} %$\cap$ \{$|\Kc_{v'}| = |\Kc_{\tilde v}|: \forall \tilde v \in \Ic$\}}
\STATE $\Ic = \Ic \cup v'$; 
\ENDIF
\ENDFOR
\STATE Color all the vertices in $\Ic$ by $c \notin \Cc$; \\
\STATE Let $\cv[\Ic] = c$; \\
%\STATE $\hat\Vc = \hat\Vc \setminus \Ic$;
%\STATE Color all the vertices in $\Ic$ by an unused color.
\STATE $\Vc = \Vc \setminus \Ic$.
\ENDWHILE
\RETURN $\max_{v\in\mathcal{V}}|\cv(\mathcal{N}^+(v))|$ and the corresponding $\cv(\mathcal{N}^+(v))$ for each $v$;
\end{algorithmic}
}
\end{algorithm}

\begin{algorithm}[ht]
\caption{HgLC$_1$}
\label{algorithm: HgLC 1}
\small{
\begin{algorithmic}[1]
\STATE $\Cc = \emptyset$;
\STATE $\cv = \emptyset$;
\STATE choose  $a \in [0, 1]$ 
\STATE choose  $b \in [0, 1]$ 
%\STATE $\Gc_{n+1} = \emptyset$
%\STATE $\Gc_{0} = \emptyset$
\FORALL{$i = n, n-1, \dots, 2,1$} 
%\STATE $\Gc_i = \Gc_{i+1} \cup \Gc_{i}$
\FORALL{$v \in \Gc_i$ and $|\Kc_v| = i$}
\STATE $\Ic = \{v\}$;
\FORALL{$v' \in \Gc_i \setminus \Ic$ with $|\Kc_{v'}| = |\Kc_{v}|$ }
%\IF{\{There is no edge between $v'$ and $\Ic$\} $\cap$ \{$|\Kc_{v'}| = |\Kc_{\tilde v}|: \forall \tilde v \in \Ic$\} } 
\IF{\{There is no edge between $v'$ and $\Ic$\}} 
\STATE $\Ic = \Ic \cup v'$;
\ENDIF
\ENDFOR
\IF {$|\Ic| = i$}
%\STATE \%\%\%\%\%\%\%\%\%\%\%\%\%\%\%\%\%\%\%\%\%\%\%\%\%\%\%\% 
%\STATE It is also possible to put $|\Ic| \geq i-1$ at line 11 to check
%\STATE \%\%\%\%\%\%\%\%\%\%\%\%\%\%\%\%\%\%\%\%\%\%\%\%\%\%\%\%
\STATE Color all the vertices in $\Ic$ by $c \notin \Cc$;
\STATE $\cv[\Ic] = c$, $\Cc = \Cc \cup c$;
%\STATE $\Cc = \Cc \cup c$
\STATE $\Gc_i = \Gc_i \setminus \Ic$;
%\ELSE
%\STATE Go to line 25 %21
\ENDIF
\ENDFOR
%\STATE $\Ic = \emptyset$
%\STATE Let $\gv_i$ be the vector representing the vertices in $\Gc_i$, where $\gv_i[j]$ is $v \in \Gc_i$ with the $j$th smallest $|\Kc_v|$. 
%\STATE Let $\Wc = \{v \in \Gc_i: |\Kc_v| \geq a \cdot \min_{v \in \Gc_i} |\Kc_v|\}$, where $a$ is a parameter to be decided. 
%\FORALL {$v \in \Gc_i$ with smallest $|\Kc_v|$} 
\FORALL{$v \in \Gc_i$ with $v$ randomly picked from $\Wc_1 \subset \Gc_i$}%{Randomly pick a $v \in \Wc_1 \subset \Gc_i$} %with \\$\Wc_1 = \{v \in \Gc_i: \min_{v \in \Gc_i} |\Kc_v|\leq |\Kc_v| \leq  \min_{v \in \Gc_i} |\Kc_v|  + \lfloor a \left( \max_{v \in \Gc_i} |\Kc_v| -  \min_{v \in \Gc_i} |\Kc_v|\right) \rfloor\}$}%with smallest $|\Kc_v|$} 
%\FORALL {$j = 1, \cdots$ with smallest $|\Kc_v|$} 
\STATE $\Ic = \{v\}$;
\STATE $\Qc_i = \Gc_i \setminus \Ic$;
%\FORALL{$v' \in \Gc_i \setminus \Ic$ with smallest $|\Kc_{v'}|$}
%\FORALL{$v' \in \Qc_i$ with smallest $|\Kc_{v'}|$}
\FORALL{$v' \in \Qc_i$ with $v'$ randomly picked from $\Wc_2 \subset \Qc_i$.} %, with $\Wc_2 = \{v' \in \Qc_i: \min_{{v'} \in \Qc_i} |\Kc_{v'}|\leq |\Kc_{v'}| \leq \min_{{v'} \in \Qc_i}  |\Kc_{v'}| + \lfloor b \left(\max_{{v'} \in \Qc_i}  |\Kc_{v'}| - \min_{{v'} \in \Qc_i}  |\Kc_{v'}| \right)  \rfloor\} $ } %with smallest $|\Kc_{v'}|$}
%\FORALL{$v' \in \Gc_i/\{v\}$}
\IF{\{There is no edge between $v'$ and $\Ic$\}}
\STATE $\Ic = \Ic \cup v'$;
\STATE $\Qc_i = \Qc_i \setminus \{v'\}$; %$\Qc_i = \Qc_i \setminus \Ic$
\ELSE
\STATE $\Qc_i = \Qc_i \setminus \{v'\}$;
\ENDIF
\ENDFOR
\IF{$|\Ic| \geq i$}
\STATE Color all the vertices in $\Ic$ by $c \notin \Cc$;
\STATE $\cv[\Ic] = c$, $\Cc = \Cc \cup c$;
%\STATE $\Cc = \Cc \cup c$
\STATE $\Gc_i = \Gc_i \setminus \Ic$;
\ELSE
%\STATE Go to line 4
\STATE $\Gc_i = \Gc_i \setminus \{v\}$, $\Gc_{i-1} = \Gc_{i-1} \cup \{v\}$;
%\STATE $\Gc_{i-1} = \Gc_{i-1} \cup \{v\}$
\ENDIF
\ENDFOR
\ENDFOR
\STATE $\cv=$LocalSearch($\Hc_{\Cm, \Wm}, \cv, \Cc$); 
\RETURN $\max_{v\in\mathcal{V}}|\cv(\mathcal{N}^+(v))|$ and the corresponding $\cv(\mathcal{N}^+(v))$ for each $v$; 
\end{algorithmic}
}
%\vspace{-0.5cm}
\end{algorithm}
%\vspace{-0.5cm}

%\begin{figure}[th!]
%\centerline{\fbox{\begin{minipage}[h]{7.5 cm} \normalsize{ %11.8
%{\bf algorithm} {\tt GCLC$_1$} \\
%$1$\hspace*{0.2truecm} Let $\hat \Vc = \Vc$; \\
%$2$\hspace*{0.2truecm} Let $\Cc = \emptyset$; \\
%$3$\hspace*{0.2truecm} $\cv_{\rm 1} = \emptyset$; \\
%$4$\hspace*{0.2truecm} {\bf while} $\hat \Vc \neq \emptyset$ \\
%$5$\hspace*{0.7truecm} Pick an arbitrary vertex $v$ in $\hat \Vc$. Let $\Ic = \{v\}$; \\
%$6$\hspace*{0.7truecm} {\bf for} all $v' \in \hat \Vc/\{v\}$$\rightarrow$ \\
%$7$\hspace*{1truecm} {\bf if} (There is no edge between $v'$ and $\Ic$\} \\
%\hspace*{1.2truecm} $\cap$ \{$\Kc_{v'} = \Kc_{\tilde v}: \forall \tilde v \in \Ic$) {\bf then}  \\
%$8$\hspace*{1.2truecm} $\Ic = \Ic \cup v'$; \\
%$9$\hspace*{1truecm} {\bf endif}\\
%$10$\hspace*{0.5truecm} {\bf endfor} \\
%$11$\hspace*{0.5truecm} Color all the vertices in $\Ic$ by $c \notin \Cc$; \\
%$12$\hspace*{0.5truecm} Let $\cv_1[\Ic] = c$; \\
%$13$\hspace*{0.5truecm} $\hat\Vc = \hat\Vc \setminus \Ic$; \\
%$14$\hspace*{0.0truecm} {\bf endwhile}                            \\
%$15$\hspace*{0.0truecm} {\bf return}($\cv_{\rm 1}$);          \\
%{\bf end} {\tt GCLC$_1$} }
%\end{minipage}}}
%\caption{The greedy constrained coloring algorithm GCLC$_1$ achieving (\ref{eq: psi}) for large enough $F$. $\Kc_v$ denotes the set of users that are either caching or requesting packet $v$.}
%\label{algorithm: coloring 1}
%\end{figure}
%Let the $\Kc_v$ denote the set of users that are either caching or requesting packet $v$. %(see Example \ref{example: algorithm 1}). 

On the other hand, GCLC$_2$ computes the minimum coloring of $\Hc_{\Cm,\Wm}$ subject to the constraint that only the vertices representing the same packet are allowed to have the same color. In this case, the total number of colors is equal to the number of distinct requested packets, and the coloring can be found in $O(|\Vc|^2)$. Then, it remains to find $\max_{v\in\mathcal{V}}|\cv(\mathcal{N}^+(v))|$. It can be seen that this scheme achieves the same rate as sending linear random combinations of all  requested packets. %corresponds to the naive (uncoded) multicasting transmission of all requested packets.  %In another word, we can just transmit this packet such that all the users can receive (decode) simultaneously.

This scheme, GCLC, was shown in \cite{ji2014order2} to be order-optimal when $B \rightarrow \infty$ for the homogeneous shared link network with $L$ requests per user, in the worst-case demand setting. %is considered instead of the average demand.

%In Appendix \ref{sec: Proof of Corollary}, we prove that GCLC achieves the upper bound of the average rate for the homogeneous shared link networks given by Corollary \ref{corollary: result homogeneous} (when $B \rightarrow \infty$ such that $F \rightarrow \infty$). However, as will be shown in Section \ref{sec: Simulations and Discussions}, for finite $F$, GCLC loses the promising multiplicative caching gain.

\subsection{Hierarchical greedy Local Coloring (HgLC)}

Similar to GCLC, HgLC also works by first computing two valid local colorings of the conflict graph $\Hc_{\Cm,\Wm}$, referred to as HgLC$_1$ and HgLC$_2$. Then, the transmission code is constructed based on the coloring with minimum rate. In this case, HgLC$_2$ is the same as GCLC$_2$, while HgLC$_1$ is described by Algorithm \ref{algorithm: HgLC 1}.

%In Algorithm \ref{algorithm: HgLC 1}, %let $g(v)$ be the packet represented by vertex $v$, we defined $\Kc_v = \{\forall u \in \Uc: g(v) \in \Qm_{u} \cup \Cm_u\}$, where $\Qm_u$ is the set of all the requested packets by user $u$ and $\Cm_u$ is the set of all the cached packets by user $u$. 
We guide the reader through Algorithm \ref{algorithm: HgLC 1} in the following. 
Let  $\Kc_v = \{\forall u \in \Uc: \rho(v) \in \Wm_{u} \cup \Cm_u\}$ and
 $\Gc_i = \{v: |\Kc_v| = i\}$. We consider $\Gc_i$ to represent the $i$th hierarchy. For HgLC$_1$, we start from hierarchy $n$. First, we color a subset of vertices in $\Gc_n$ with the same color if: they have the  same $|\Kc_v|=n$, the cardinality of such set is $n$, and there are no links between any two vertices of such set in the conflict graph.  Then, we merge the uncolored vertices in $\Gc_n$ with $\Gc_{n-1}$ ($\Gc_{n-1} = \Gc_{n-1} \cup \Gc_{n}$, line 35 of Algorithm \ref{algorithm: HgLC 1}) to form a new hierarchy $n-1$. In the hierarchy $n-1$, again, we first color  a subset of vertices in $\Gc_{n-1}$  with the same color if: they have the  same $|\Kc_v|=n-1$, the cardinality of such set is $n-1$ and there are no links between any two vertices of such set in the conflict graph.  Next, we try to color the uncolored vertices in $\Gc_{n-1}$ according to the following procedure:  1) randomly pick a vertex $v$ from $\Wc_1 \subset \Gc_{n-1}$   , 2) color with the same color the chosen vertex $v$ and the other vertices $v' \in \Wc_2$ whose $|\Kc_{v'}|$ are ``close'' to $|\Kc_{v}|$ in a greedy manner.
Here,  $\Wc_1$ denotes a set of vertices with ``small'' $|\Kc_v|, v \in \Gc_{n-1}$ or ``large degree'' in $\Hc_{\Cm,\Wm}$ and the value of $a \in [0,1]$ control the size of $\Wc_1$. Formally, $\Wc_1 = \{v \in \Gc_i: \min_{v \in \Gc_i} |\Kc_v|\leq |\Kc_v| \leq  \min_{v \in \Gc_i} |\Kc_v|  + \lfloor a \left( \max_{v \in \Gc_i} |\Kc_v| -  \min_{v \in \Gc_i} |\Kc_v|\right) \rfloor\}$. For example, if $a=0$, then $\Wc_1$ denotes the vertex with the smallest $|\Kc_v|$. 
$\Wc_2 \subset \Gc_i \setminus \{v\}$  is defined as  $\Wc_2 = \{v' \in \Qc_i: \min_{{v'} \in \Qc_i} |\Kc_{v'}|\leq |\Kc_{v'}| \leq \min_{{v'} \in \Qc_i}  |\Kc_{v'}| + \lfloor b \left(\max_{{v'} \in \Qc_i}  |\Kc_{v'}| - \min_{{v'} \in \Qc_i}  |\Kc_{v'}| \right)  \rfloor\}$, where $\Qc_i$ is defined in Algorithm \ref{algorithm: HgLC 1} and  $b \in [0,1]$. For example, if $b=0$, then we start from the vertex $v'$ such that $|\Kc_{v'}| - |\Kc_{v}|$ is minimized. Here, we are looking for the independent set with size at least $i$ in the $i$th hierarchy in a greedy manner. After this second coloring procedure, we union the uncolored vertices with the vertices of the next hierarchy, which in this case, is $\Gc_{n-2}$. Then, we repeat the same procedure for all the hierarchies. Finally,  in line 39 of Algorithm \ref{algorithm: HgLC 1}, we use a function called LocalSearch to further reduce the required number of colors. 
The LocalSearch function is described by Algorithm \ref{algorithm: Local_Search}
%For example, if vertex $v_1$ representing packet $A_1$ ($f(v_1) = A_1$) is requested by user $1$ and user $2$, and cached in user $3$ and user $4$, then $\Kc_{v_1} = \{1 \,\, 2, 3 \,\, 4\}$. For clarity, we use a comma to separate the user requesting $A_1$ and the users caching $A_1$.
%$\cv$ is the desired vector showing the coloring. 
%A detailed description of HgLC$_1$ is given by Algorithm \ref{algorithm: HgLC 1}, where $\Nc(j)$ denote the neighbors of vertex $j$ excluding vertex $j$.
%On the other hand, $HgLC$_2$ is the same as GCLC$_2$. 
It can be shown that the complexity of HgLC$_1$ is given by $O(n|\Vc|^2)$. %$O(n^3B^2)$. %In the following, we will give an example to describe HgLC$_1$. 

\begin{algorithm}[ht]
\caption{LocalSearch($\Hc_{\Cm, \Wm}, \cv, \Cc$)}
\label{algorithm: Local_Search}
{\small
\begin{algorithmic}[1]
\FORALL{$c \in \Cc$}
\STATE Let $\Jc_c$ be the set of vertices whose color is $c$; 
\STATE Let $\Bc=\emptyset$;
\STATE Let $\hat \cv= \cv$;
\FORALL{$i \in \Jc_c$}
\STATE $\Ac = \emptyset$; 
\FORALL{$j \in \Nc(i)$}
\STATE $\Ac = \Ac \cup \cv[j]$; 
\IF{$\Cc \setminus \Ac \neq \emptyset$}
\STATE $c'$ is randomly picked from $\Cc \setminus \Ac$; 
\STATE $\hat \cv[i] = c'$;
\STATE $\Bc=\Bc\cup \{i\}$;
\ENDIF
\ENDFOR
\IF{$|\Bc| = |\Jc_c|$}
\STATE $\cv = \hat \cv$;
\STATE $\Cc = \Cc \setminus c$; 
\ENDIF
\ENDFOR
\ENDFOR
\RETURN $\cv$;
\end{algorithmic}
}
\end{algorithm}

%%%%%%%%%%%%%%%%%%%%%%%%%%%%%%%%%%%%%%%%%%%%%%%%%%%%%%%%%%
%\vspace{-0.5cm}
\section{Simulations and Discussions}
\label{sec: Simulations and Discussions}

%As already mentioned in the previous subsections in  \cite{ji2014average},  \cite{maddah2012fundamental}, and \cite{maddah2013decentralized}  schemes exploiting the combined benefit of network coding and in-network cooperative caching capabilities have been proposed and shown to be able to achieve significant gains in caching network with a shared multicast link under the assumption of infinite packetizations. 
In this section, we numerically analyze the performance of HgLC %the two achievable schemes illustrated in Section \ref{sec: algorithms} 
for finite file packetization by assuming the  random popularity-based (RAP) caching policy in Algorithm \ref{alg1}. 
We first introduce the achievable rate of the benchmark scheme LFU (Least Frequently Used)\footnote{LFU discards the least frequently requested file upon the arrival of a new file to a full cache of size $M_u$ files. In the long run, this is equivalent to caching the $M_u$ most popular files.}, 
given by:
%. By using the second term of (\ref{eq:2}) given by Theorem \ref{thm:up}, we have the achievable rate for LFU as
%\begin{theorem}
%The average rate of LFU is given by
\be
R^{\rm LFU} = \sum_{f = \min_u\{M_u\} + 1}^m \left(1 - \prod_{u \in \Uc_{\{M_u < f\}}} (1-q_{f,u})^{L_u}\right),
\ee
where $\Uc_{\{M_u < f\}}$ denotes the set of users with $M_u < f$. 
%\hfill  $\square$
%\end{theorem}

%For the homogeneous case, we have the following corollary. 
%\begin{corollary}
%Let $q_{f,u} = q_f, \forall f \in \Fc$, $M_u = M, \forall u \in \Uc$ and $L_u = L, \forall u \in \Uc$, then the average rate of LFU-NM is given by
%\be
%\bar R^{\rm LFU} = \sum_{f = M + 1} \left(1 - (1-q_{f})^{nL}\right).
%\ee
%\hfill  $\square$
%\end{corollary}

%In this section, we numerically analyze the performance of the two achievable schemes illustrated in Section \ref{sec: algorithms} for finite file packetization. Specifically, assuming the distributed random popularity-based caching policy in Fig. \ref{alg1}, we compare the average performance of GCC and GRASP when files are partitioned into a finite number ($B$) of packets.  
%For comparison, we also plot 1) the performance of LFU (Least Frequently Used),\footnote{LFU discards the least frequently requested file upon the arrival of a new file to a full cache of size $M_u$ files. In the long run, this is equivalent to caching the $M_u$ most popular files.}  shown to be optimal in uncoded networksand  2) the performance of GCC for infinite packetization ($B\rightarrow \infty$), as given in Theorem \ref{thm:up}.

For simplicity and to illustrate the effectiveness of HgLC, we consider a homogenous scenario in which
users request files according to a common Zipf demand distribution with parameter $\gamma \in \{0.2, 0.4\}$ and all caches have size $M$ files. We assume two types of users. In one case, they represent end devices requesting only one file each ($L=1$). In a second case, they  represent helpers/small-cells, each serving $10$ end user devices, and consequently collecting at most $L=10$ distinct requests. 
Moreover, we let the caching distribution to be uniform, which means that $\Pm$ is chosen as a $m$-dimensional vector taking value of $\frac{1}{m}$.\footnote{The caching distribution $\Pm^*$  can be obtained by minimizing $ R^{\rm GCLC}(\Pm,\Qm)$ in (\ref{eq:2}) among all $\Pm$ described by a $m$-dimensional  vector taking  value in $ \{\frac{1}{\widetilde{m}}, 0\}$ in practice, as suggested in \cite{ji2014average}. } %This constraint on the caching distribution introduced in \cite{ji2014average},  originates a scheme referred to as Random Least-Frequently-Used (Random LFU), which approximates RAP and generalizes the well known LFU caching scheme.
%In Random LFU, each user just caches packets from the (carefully designed) $\tilde{m}$ most popular files in a distributed and random manner.}

%For all considered schemes, the caching distribution $\Pm^*$  is obtained by minimizing $\bar R^{\rm ub}(\Pm,\Qm)$ in (\ref{eq:2}) among all $\Pm$ described by a $m$-dimensional  vector taking  value in $ \{\frac{1}{\tilde{m}}, 0\}$.\footnote{This constraint on the caching distribution introduced in \cite{ji2014average},  originates a scheme referred to as Random Least-Frequently-Used (Random LFU), which approximates RAP and generalizes the well known LFU caching scheme.
%In Random LFU, each user just caches packets from the (carefully designed) $\tilde{m}$ most popular files in a distributed and random manner.},
%%{\BLUE {\RED We refer GCC as the two algorithm achieving the upper bound before, so GCC definitely includes LFU. We may want to rewrite this BLUE sentences.} 

%Further, we assume that when using GCC or GRASP, %based on the system parameters (number of users, $n$, number of files, $m$, number of packets, $B$ and Zip parameter $\gamma$),  
%the source node pre-evaluates the performance of LFU and chooses the minimum accordingly. %to adopt between LFU and  GCC, the scheme minimizing the result rate over the shared link.  Similarly assumption is done for GRASP. 
%Hence, denoting by $R_{LFU}$, $R_{GCC}$ and $R_{GRASP}$ the average rate achieved by LFU, GCC and GRASP, respectively, Fig. \ref{fig: result 1} plots the performance of GCC and GRASP as  $\min\{R_{LFU}, R_{GCC}\}$, and $\min\{R_{LFU}, R_{GRASP}\}$, respectively.

\begin{figure}[ht]
%\vspace{-3cm}
\centering
\hspace{-0.7cm}
\subfigure[]{
%\vspace{-5cm}
%\hspace{-1cm}
\centering \includegraphics[width=4.7cm, height=4cm]{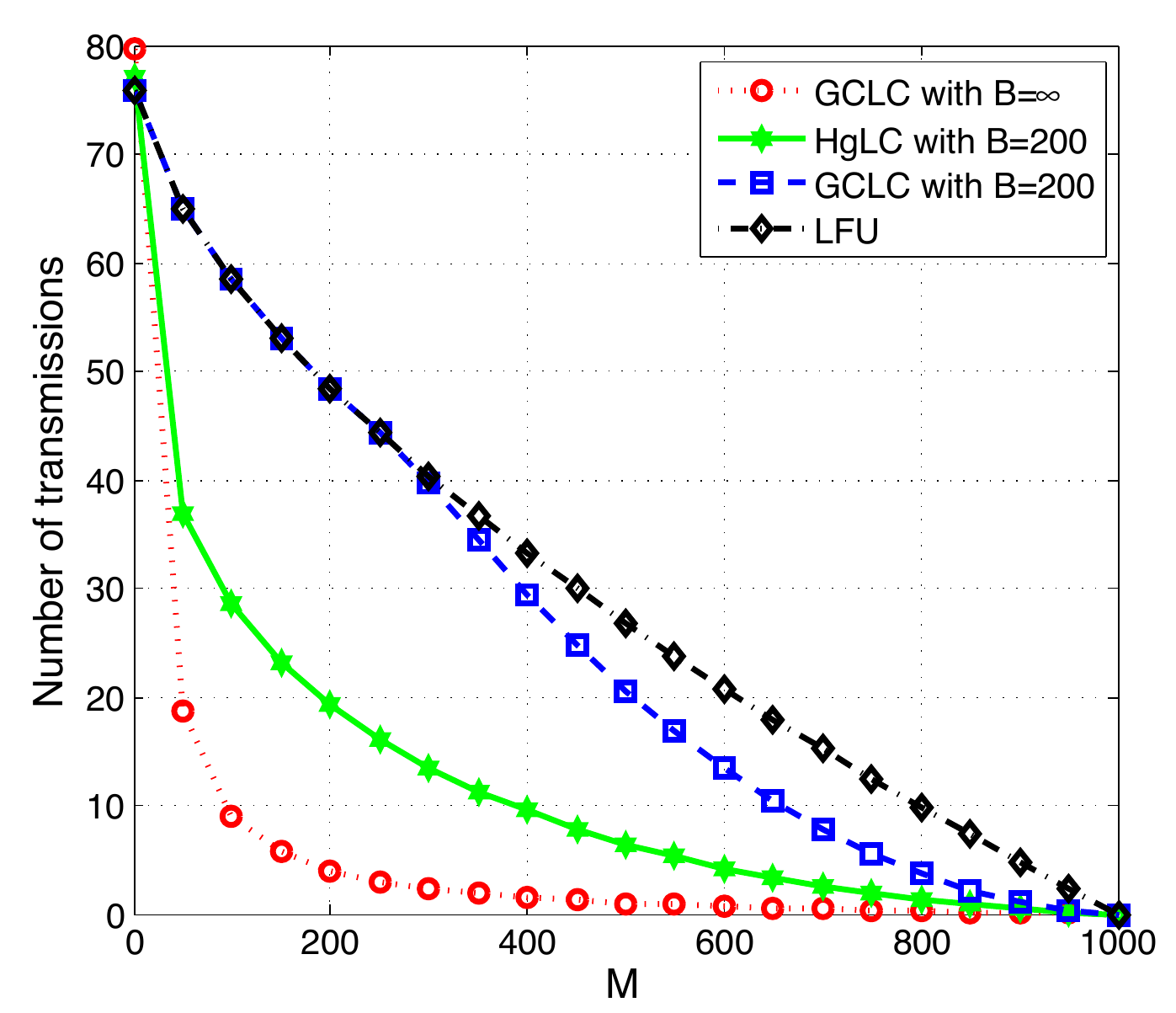}
%\vspace{-5cm}
%\hspace{-3cm}
\label{fig1}
}
\hspace{-0.7cm}
\subfigure[]{
%\vspace{-5cm}
%\hspace{-1cm}
\centering \includegraphics[width=4.7cm, height=4cm]{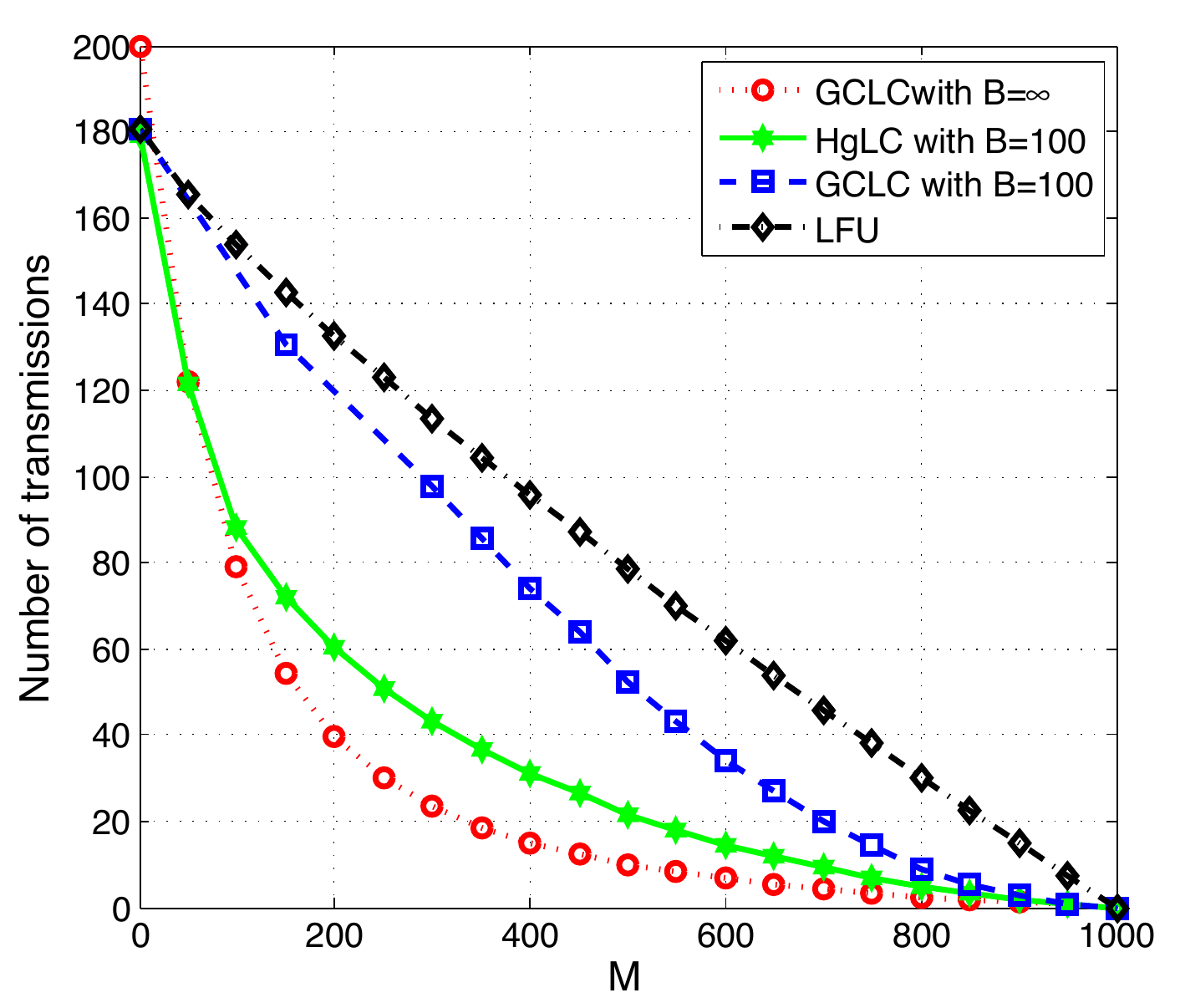}
%\vspace{-5cm}
%\hspace{-3cm}
\label{fig2}
}
%\hspace{-0.8cm}
%\subfigure[]{
%%\vspace{-5cm}
%%\hspace{-0.8cm}
%\centering \includegraphics[width=4.8cm, height=4cm]{n20m100b100gamma02n15n215L10}
%%\vspace{-5cm}
%%\hspace{-3cm}
%\label{fig3}
%}
%\hspace{-0.8cm}
%\subfigure[]{
%%\vspace{-5cm}
%%\hspace{-1cm}
%\centering \includegraphics[width=4.8cm, height=4cm]{n20m100b100gamma02n15n215L20}
%%\vspace{-5cm}
%%\hspace{-3cm}
%\label{fig4}
%}
\vspace{-0.5cm}
\caption{Average number of transmissions in a heterogeneous shared link network with $m=1000$. a)~$n=80$, $L=1$, $\alpha=0.4$; b)~$n=20$, $L=10$, $\alpha=0.2$. %Infinite File Length indicate the rate of GCLC when $B \rightarrow \infty$ given in Theorem \ref{thm:up}. 
}
\label{fig: result 1}
\end{figure}

Fig. \ref{fig1} plots the average rate for a network with $n=80$ users with $\gamma =0.4$, when $L=1$ and $B=200$. 
  %consisting of $n_1=5$ and $n_2=15$, $L_1 \in \{1,5,10,20\}$, $L_2=1$,  $m=100$ files and $B=100$ packets. %and Zipf parameter $\gamma=0.2$. 
Observe how the significant caching gains (with respect to LFU) quantified by the upper bound (GCLC with $B=\infty$) are completely lost when using GCLC with finite packetization. %and only slightly recovered as the packetization increases to $B=100$ and $B=200$. 
On the other hand, observe how HgLC remarkably preserves most of the promising multiplicative caching gains for the same values of file packetization. 
For example, in Fig.~\ref{fig1}, if $M$ doubles from $M=200$ to $M=400$, then the rate achieved by HgLC essentially halves from $20$ to $10$. Furthermore, HgLC is able to achieve a factor of $5$ rate reduction from LFU for $M=500$.
For the same regime, it is straightforward to verify that neither GCLC nor LFU exhibit this property.\footnote{While LFU can only provide an additive caching gain, additive and multiplicative gains may show indistinguishable when $M$ is comparable to the library size $m$. Hence, one needs to pick a reasonably small $M$ ($\frac{m}{n} < M \ll m$) to observe the multiplicative caching gain of HgLC.} 
%Fig. \ref{fig: result 1}(d) illustrates a scenario with higher popularity skewness, e.g., $\gamma=0.6$. Observe how, also in this scenario, a finite number of packets ($B=50$) completely limits the gains of GCLC. On the other hand, HgLC is still able to preserve significant gains. For example, when $M$ doubles from $M=70$ to $M=140$, the achievable rate by HgLC goes from $8.8$ to $5.5$, approaching a half rate reduction even with only $50$ packets per file.  
Note from Fig. ~\ref{fig1}, that in order to guarantee a rate of $20$, GCLC requires a cache size of $M=500$, while HgLC can reduce the cache size requirement to $M=200$, a $2.5\times$ cache size reduction. 
Finally, we notice that the computational time required by HgLC in the scenario of Fig. \ref{fig1}, with $B=200$ and $M=200$ (20\% of the library size), computed as Matlab-cputime $/10$ on an intel i5 2.4 GHz processor, %using as process Intel i$5$ $2.4$ GHz, 
is around $30s$.
%In practice, this would translate, for example, in a delay of $30s$ before users start streaming video files partitioned into segments of duration at least $30s$ each.

Fig.~\ref{fig2},  plots the average rate for a network with $n=20$ helper/small-cell each serving $10$ users 
making requests according to a Zip distribution with  $\gamma=0.2$. Hence the total number of distinct requests per helper is up to $L_u = 10, \forall u \in \{1,\ldots,20\}$. In this case, we assume $B=100$.
Observe first the order-optimal asymptotic rate (shown in red). Note from Figs.  \ref{fig1} and \ref{fig2} that when $L_u$ increases (from $L_u=1$ to $L_u=10$), while the average rate per request reduces, the gains with respect to LFU also reduce. This is explained by the fact that when aggregating multiple requests per user, there is a higher number of overlapping requests, which increases the opportunities for naive multicasting, as clearly characterized in \cite{ji2014order2}). Note, however, that HgLC is able to, remarkably, keep similar gains with respect LFU in this multiple request setting, and approach the asymptotic performance even with just $B=100$ packets per file, confirming the effectiveness of the local coloring procedures in HgLC.

\bibliographystyle{IEEEbib}
\bibliography{references,references_d2d}

\end{document}

%% file: macros.tex
\long\def\comment#1{}

\newfont{\bbb}{msbm10 scaled 700}

\newfont{\bb}{msbm10 scaled 1100}

\newcommand{\PP}{\mbox{\bb P}}

\newcommand{\FF}{\mbox{\bb F}}

\newcommand{\EE}{\mbox{\bb E}}

\newcommand{\cv}{{\bf c}}

\newcommand{\fv}{{\bf f}}
\newcommand{\gv}{{\bf g}}

\newcommand{\qv}{{\bf q}}

\newcommand{\Cm}{{\bf C}}

\newcommand{\Gm}{{\bf G}}

\newcommand{\Pm}{{\bf P}}
\newcommand{\Qm}{{\bf Q}}

\newcommand{\Wm}{{\bf W}}

\newcommand{\Ac}{{\cal A}}
\newcommand{\Bc}{{\cal B}}
\newcommand{\Cc}{{\cal C}}
\newcommand{\Dc}{{\cal D}}
\newcommand{\Ec}{{\cal E}}
\newcommand{\Fc}{{\cal F}}
\newcommand{\Gc}{{\cal G}}
\newcommand{\Hc}{{\cal H}}
\newcommand{\Ic}{{\cal I}}
\newcommand{\Jc}{{\cal J}}
\newcommand{\Kc}{{\cal K}}

\newcommand{\Nc}{{\cal N}}

\newcommand{\Qc}{{\cal Q}}

\newcommand{\Tc}{{\cal T}}
\newcommand{\Uc}{{\cal U}}
\newcommand{\Wc}{{\cal W}}
\newcommand{\Vc}{{\cal V}}

\newcommand{\omegav}{\hbox{\boldmath$\omega$}}

\newcommand{\Csf}{{\sf C}}

\newcommand{\Hsf}{{\sf H}}

\newcommand{\Wsf}{{\sf W}}

\renewcommand{\arg}{{\hbox{arg}}}

\newcommand{\eqdef}{\stackrel{\Delta}{=}}

\newcommand{\transp}{{\sf T}}

\newcommand{\be}{\begin{equation}}
\newcommand{\ee}{\end{equation}}
\newcommand{\bea}{\begin{eqnarray}}
\newcommand{\eea}{\end{eqnarray}}